\documentclass[11pt, a4paper]{article}

\usepackage{fullpage}

\usepackage{microtype}

\title{A Short Note on Two-Variable Logic with a Linear Order Successor and a Preorder Successor}

\author{Amaldev Manuel\\
\small LIAFA, Universit\'e Paris Diderot\\ 
\small  \texttt{amal@liafa.univ-paris-diderot.fr}
\and 
Thomas Schwentick\\
\small TU Dortmund University\\
\small \texttt{thomas.schwentick@cs.uni-dortmund.de}
\and 
Thomas Zeume\\
\small TU Dortmund University\\
\small  \texttt{thomas.zeume@cs.uni-dortmund.de}
}

\usepackage{	amssymb,
		amsmath,
		amsthm}

\usepackage{	xspace,
		enumerate,
		gensymb,
		xspace,
		graphicx,
		graphics,
		multirow,
		sidecap,
		enumerate}

\usepackage{	datetime}

\usepackage{	color, 
		colortbl, 
		hhline}

\usepackage{	arydshln}
\usepackage{	pst-node,
		pst-plot,
		pst-pdf}

\usepackage{	rotating}
\usepackage{	tikz}

\usepackage{hyperref}
\usepackage{ifmtarg}
\usepackage{chngcntr}

\clearpage{}\makeatletter{}

\newcommand{\myemph}[1]{\emph{#1}}

\theoremstyle{plain}

\newtheorem{theorem}{Theorem}

\newenvironment{proofof} [1]{\noindent{\bf Proof (of #1).}\enspace}{\qed  \vspace{2mm}}

\makeatletter
\newcommand{\theoremcont}[3]{
   \def\Type{#1}
   \def\Number{#2}
   \def\Label{#3}
   \@ifmtarg{#3}{
     \vspace{2mm}\textbf{\Type\ \Number.}\itshape\noindent
   }{
    \vspace{2mm}\textbf{\Type\ \Number\ (\Label).}\itshape\noindent
  }
}

\newcommand{\fotwo}{\ensuremath{\mathrm{FO}^2}\xspace}

\newcommand{\tpo}{total preorder\xspace}

\newcommand{\tlo}{linear order\xspace}

\newcommand{\po}[1][]{\ensuremath{\leq_{p_{#1}}}}
\newcommand{\spo}[1][]{\ensuremath{<_{p_{#1}}}}
\newcommand{\lo}[1][]{\ensuremath{\leq_{l_{#1}}}}
\newcommand{\slo}[1][]{\ensuremath{<_{l_{#1}}}}
\newcommand{\psim}[1][]{\ensuremath{\sim_{p_{#1}}}}

\newcommand{\psucc}[1][]{\ensuremath{{+1}_{p_{#1}}}}

\newcommand{\lsucc}[1][]{\ensuremath{{+1_{l_{#1}}}}}

\newcommand {\calA}      {{\cal A}\xspace}

\newcommand {\calM}      {{\cal M}\xspace}

\newcommand  {\N}   {\ensuremath{\mathbb{N}}}

\newcommand{\inc}{\ensuremath{\text{inc}}}
\newcommand{\dec}{\ensuremath{\text{dec}}}
\newcommand{\ifzero}{\ensuremath{\text{ifzero}}}
\newcommand{\ifz}{\ifzero}

\newcommand*\colvec[3][]{
    \begin{pmatrix}\ifx\relax#1\relax\else#1\\\fi#2\\#3\end{pmatrix}
}
\clearpage{}

\begin{document}

\maketitle

\begin{abstract}
  The finite satisfiability problem of two-variable logic extended by a linear order successor and a preorder successor is shown to be undecidable.
\end{abstract}

  \makeatletter{}The decidability of the finite satisfiability problem of two-variable logic extended by orders and preorders  as well as their corresponding successor relations has been currently investigated by several papers \cite{Manuel10, SchwentickZ11, ManuelZ13}. This short note extends \cite{ManuelZ13} which is work under submission. In the setting of \cite{ManuelZ13} only two cases remained open. One of them, namely the extension of two-variable logic by one successor relation of a linear order and one successor relation of a preorder, is settled here. We refer to \cite{Manuel10} and \cite{SchwentickZ11} for more background and motivation.

In the rest of this note we first introduce the necessary notation and then prove that the finite satisfiability problem of the extension of two-variable logic by one successor relation of a linear order and one successor relation of a preorder is undecidable.

\section{Notations}
A binary relation $\po$ over a finite set\footnote{In this note all sets are finite.} $A$ is a \myemph{preorder} if it is reflexive, transitive and total, that is, if for all elements $u$,$v$ and $w$ from $A$ (i) $u \po u$ (ii) $u \po v$ and $v \po w$ implies $u \po w$ and (iii) $u \po v $ or  $v \po u$ holds. A \myemph{\tlo} $\lo$ on $A$ is an antisymmetric total preorder, that is, if $u \lo v$ and $v \lo u$ then $u = v$.

Thus, the essential difference between a \tpo and a \tlo is that the former allows for two distinct elements $u$ and $v$ that both  $u
\po v$ and $v \po u$ hold. We call two such elements \myemph{equivalent with respect to $\po$} and denote this by $u \sim_p v$.
Hence, a \tpo can be seen as an equivalence relation $\psim$ whose equivalence classes are linearly ordered by a linear order. Clearly, every \tlo is a \tpo with equivalence classes of size one. We write  $u \slo v$ if $u \lo v$ but not $v \lo u$, analogously for a preorder order $\po$. Further, if $C$ and $C'$ are the equivalence classes of $u$ and $v$, respectively, then we write $C \po C'$ if $u \po v$.

For a linear order $\lo$ an induced \myemph{successor relation} $\lsucc$ can be defined in the usual way, namely by letting $\lsucc(u,v)$ if and only if $u \slo v$ and there is no $w$ with $u \slo w \slo v$. Similarly a preorder $\po$ induces a successor relation $\psucc$ based on the linear order on its equivalence classes, i.e. $\psucc(u,v)$ if  and only if $u \spo v$ and there is no $w$ with $u \spo w \spo v$. Thus an element can have several successor elements in $\psucc$.

An \myemph{ordered structure} is a finite structure with non-empty universe and some linear orders, some total preorders, some successor relations and some unary relations. Linear orders and their induced successor relations will be denoted by $\lo, \lo[1], \lo[2],\ldots$ and $\lsucc, \lsucc[1], \lsucc[2], \ldots$, respectively. Analogously, preorders  and their induced successor relations will be denoted by $\po, \po[1], \po[2],\ldots$ and $\psucc, \psucc[1], \psucc[2], \ldots$, respectively. For a set of binary relation symbols $O$, an $O$-structure is a finite structure with some unary relations and some binary relations of types indicated by $O$. For example an $(\lsucc, \psucc)$-structure is a structure with some unary relations and, following the conventions from above, a linear order successor and a preorder successor.

\myemph{Two-variable logic} $\fotwo$ is the restriction of first-order logic to formulas with at most two distinct variable $x$ and $y$. By $\fotwo(O)$ we denote two-variable logic over a vocabulary that contains some unary relation symbols and binary relation symbols from $O$ which have to be interpreted by $O$-structures. For example, formulas in $\fotwo(\lsucc, \psucc)$ can use some unary relation symbols and the binary relation symbols $\lsucc$ and $\psucc$, where $\lsucc$  and $\psucc$ have to be interpreted as a linear successor and a preorder successor.

  \makeatletter{}\section{Result}

\makeatletter{}\newcommand{\theoremlsps}{
  \begin{theorem} \label{prop:lsps}
    Finite satisfiability of $\fotwo (\lsucc, \psucc)$ is undecidable. 
  \end{theorem}
}

\newcommand{\prooftextlsps}{
  We use the following notions. An element $v$ is a \myemph{$(\lsucc, \psucc)$-successor} of an element $u$ if $\lsucc(u,v)$ and $\psucc(u,v)$. Observe that there does not need to be a $(\lsucc, \psucc)$-successor and that it is unique if it exists. Similarly $v$ is a \myemph{$(\lsucc, \psucc)$-predecessor} of $u$ if $\lsucc(v,u)$ and $\psucc(v,u)$. 

  The proof is by a reduction from the non-emptiness problem for Minsky counter automata with two counters only.    
  Intuitively, from a given CA $\calM$ with set $C = \{R, B\}$ of counters (we refer to $R$ and $B$ as the red and blue counter in the following), we construct a $\fotwo (\lsucc, \psucc)$-formula $\varphi_\calM$ such that for every accepting run 
$\rho$ of $\calM$ there is a $(\lsucc, \psucc)$-structure $\calA_\rho$ that satisfies $\varphi_\calM$, and such that from 
every model of $\varphi_\calM$ an accepting run of $\calM$ can be constructed. The formula $\varphi_\calM$ uses as propositions the set $\Delta$ of 
transitions of $\calM$, the propositions $\{R, B\}$. We will often say color instead of proposition.

 An accepting run $\rho=\rho_0,\ldots,\rho_n$ of $\calM$ will be encoded as a $(\lsucc, \psucc)$-structure $\calA_\rho$ as follows.  The idea is to represent configurations by equivalence classes with respect to $\sim_p$, to encode transitions by propositions and counter values by the number of red and blue elements in a class. 

We will always assume in the following that $\calM$ is non-trivial and thus $\rho$ has at least one step.
More precisely, let,   for every $i\in\N_0$, $\rho_i = (p_i, (b_i, r_i))$ with $p_i \in Q$ and $b_i, r_i \in \N$ and $\delta_i$ be the transition applied in the $i$-th step. Let $k$ be large enough such that for every configuration the value of the red counter plus the value of the blue counter is at most $k$.
The intended structure $\calA_\rho$ has exactly one $\psucc$-class $\tau_i$ for each configuration $\rho_i$. 
For every $i$, $\tau_i$ has exactly $k$ elements of which $b_i$ elements carry the proposition $B$ and $r_i$ elements carry the proposition $R$. The sets of $B$-elements and of $R$-elements are disjoint. 
The relation $\lsucc$ induces bijections between successive $\psucc$-classes with the following additional properties. As the counter values of $\calM$ can change by at most one in one step, the number of blue (or red) elements in successive classes at most differs by one. The relation $\lsucc$ can thus be chosen such that 
\begin{itemize}
\item it is a bijection between the blue elements in $\tau_i$ and
  $\tau_{i-1}$, if $\delta_i$ does not change the blue counter,
\item it is a bijection between the blue elements of $\tau_i$ and the blue
  elements of $\tau_{i+1}$ minus one, in case $\delta_i$ increments
  the blue counter, and
\item  it is a bijection between the blue elements of $\tau_i$ minus one and the blue
  elements of $\tau_{i+1}$, in case $\delta_i$ decrements
  the blue counter
\end{itemize}
Likewise for the red counter and red elements.  Furthermore, each element of $\tau_i$ carries the transition $\delta_{i}$ that yielded it from the previous configuration (with the exception of $\tau_0$).

 We now state conditions (T1-T5), (B1-B5) and (R1-R5) that hold in a structure $\calA$ if and only if it is (isomorphic to) a structure $\calA_\rho$, for some accepting run $\rho$.  The conditions (T1-T5) ensure the general structure of $\calA$, the consistency of successive transitions and the initial and final state. Conditions (B1-B5) ensure that the counter values, i.e. the number of $B$-labeled elements, are consistent with the transitions. Likewise for (R1-R5).

  \begin{itemize}
    \item[(T1)] Every element that is not in the first $\psucc$-class, carries exactly one label from $\Delta$, and all elements of a $\psucc$-class carry the same $\Delta$-label.    \item[(T2)] Elements of the second $\psucc$-class are labeled with a transition starting from the start state. 
    \item[(T3)] Elements of the last $\psucc$-class are labeled with a transition leading to a final state.  
    \item[(T4)] Transitions of successive $\psucc$-classes are consistent, i.e. if $u$, $v$ are labeled with $(p, op, q)$ and $(p', op', q')$,
and  $\psucc(u,v)$, then $q = p'$.
   \item[(T5)] Every element that is not in the last $\psucc$-class has a $(\lsucc, \psucc)$-successor and every element that is not in the first class has a $(\lsucc, \psucc)$-predecessor.
  \end{itemize}

  \begin{itemize}
    \item[(B1)] Neither there are $B$-labeled elements in the first $\psucc$-class nor in the last \psucc-class.
      \item[(B2)] If some element $u$ carries a transition from $\Delta$ that does not change the $B$-counter then $\lsucc$ induces a bijection between the $B$-labeled elements of the class of $u$ and  the $B$-labeled elements of its \psucc-predecessor class.
    \item[(B3)] If some element $u$ carries a transition from $\Delta$ that increments the $B$-counter, then 
      \begin{itemize}
	\item there is exactly one $B$-labeled element $v$ with $v \psim u$ whose \lsucc-predecessor is not $B$-labeled and
        \item $\lsucc$ induces a bijection between the set of all other $B$-labeled elements of the class of $u$ and the $B$-labeled elements of its \psucc-predecessor class.
    \end{itemize}
    \item[(B4)] If some element $u$ carries a transition from $\Delta$ that decrements the $B$-counter, then
      \begin{itemize}
	\item there is exactly one $B$-labeled element $v$ with $\psucc(v,u)$ whose \lsucc-successor is not $B$-labeled and
        \item $\lsucc$ induces a bijection between the set of all other $B$-labeled elements of the $\psucc$ predecessor class of $u$ and the $B$-labeled elements of $u$'s class.
    \end{itemize}
      \item[(B5)] If some element $u$ carries a transition that tests whether the $B$-counter is equal to zero, then there is no $B$-labeled element $v$ with $\psucc(v,u)$.
  \end{itemize}

(R1)-(R5) are as (B1)-(B5) with ``blue'' replaced by ``red'', throughout. 

  For proving the correctness of this construction, we first observe that the $(\lsucc, \psucc)$-structure $\calA_\rho$ constructed from a given accepting run $\rho$ as explained above, clearly satisfies all these conditions.

  Now, let $\calA$ be a structure that satifies all conditions. Let $\tau_0, \ldots, \tau_n$ be the $\psucc$-classes of $\calA$ in increasing order. By property (T1) there is, for every $i>0$, a single transition $\delta_{i} = (p_i, op_i, q_i)$ that labels all elements from $\tau_i$. Let, for every $i$, $b_i$ and $r_i$ be the number of $B$- and $R$-labeled elements in $\tau_i$. We claim that $\rho = \rho_0, \ldots, \rho_{n}$ with $\rho_0=(p_0,(b_0,r_0))$ and, for every $i>0$, $\rho_i = (q_i, (b_i, r_i))$, is an accepting run of $M$. The states $q_i$ and $p_{i+1}$ are consistent with $\delta_i$ for all $i \in \{1, \ldots, m-1\}$; and $p_0$ is an initial state and $q_m$ is a final state, by properties (T2)-(T4).

Thus it remains to verify that the counter values in $\rho$ are consistent with the transitions. By (B1), both counters are zero in $\rho_0$. By induction over $i \in \{1, \ldots, n\}$, it can be shown that $\rho_i$ is consistent with $\rho_{i-1}$ with respect to the counter values. Condition (B2) ensures that $b_i=b_{i-1}$ if $\delta_i$ does not change the blue counter. Likewise, (B3) and (B4), respectively, ensure that  $b_i=b_{i-1}+1$ if $\delta_i$ increments the counter and  $b_i=b_{i-1}-1$ if $\delta_i$ decrements the counter. Finally, (B5) ensures that zero test transitions for the blue counter are only taken if $b_{i-1}=0$. The correctness of the values of the red counter can be proved analogously.

It remains to show that  (T1-T5), (B1-B5) and (R1-R5) can be expressed by $\fotwo (\lsucc, \psucc)$ formulas.
  
Conditions (T2)-(T5), (B1), (B2) and (B5) can be easily expressed by such formulas. 

We describe next how to construct a formula $\varphi(x)$ that expresses that in the $\psucc$-class of $x$ some element exists for which some unary formula $\psi$ holds. The idea is simply to state that for some \psucc-predecesssor of a \psucc-successor of $x$ it holds $\psi$ or  for some \psucc-successsor of a \psucc-predeccessor of $x$ it holds $\psi$. Here, we make use of our assumption that $\calA$ has at least two \psucc-classes.
That is, we define $\varphi_{\text{succ}}(x)$ as $\exists y\; (\psucc(x,y)\land \exists x (\psucc(x,y)\land \psi(x)))$. It should be stressed that the quantification might bind $x$ to some other element in the class of the ``original'' $x$. Likewise, we let $\varphi_{\text{precc}}(x)$ be $\exists y\; (\psucc(y,x)\land \exists x (\psucc(y,x)\land \psi(x)))$ and $\varphi(x)=\varphi_{\text{prec}}\lor \varphi_{\text{succ}}$. 

In this way, a formula for (T1) is readily definable. For the second part of (T1) it is stated that there is no element in the current class with some other transition proposition.

For the construction of the other formulas, we assume that (T1) and (T5) hold. 

The challenge when expressing (B3) is to make sure that there is exactly one $B$-labeled element in the current class that does not take part in the bijection induced by $\lsucc$. However, this can be expressed by stating that
\begin{enumerate}[(1)]
\item every $B$-labeled \psucc-predecessor of the current element has a $B$-labeled \lsucc-successor, and
\item there is a \psucc-predecessor of the current element $u$ that is not $B$-labeled but has a $B$-labeled \lsucc-successor, but
\item all $B$-labeled \psucc-successors of $u$ besides $u$'s \lsucc-successor have a $B$-labeled \lsucc-predecessor.
\end{enumerate}
This condition can be easily expressed by a $\fotwo (\lsucc, \psucc)$ formula.
Condition (B4) can be expressed analogously.
}

\newcommand{\prooflsps}{
  \begin{proof}
    \prooftextlsps
  \end{proof}
}

\newcommand{\proofoflsps}{
  \begin{proofof}{Theorem \ref{prop:lsps}}
    \prooftextlsps
  \end{proofof}
}

In this section we prove the only result of this note.
\theoremlsps

The proof is by a reduction from the non-emptiness problem for Minsky counter automata. Following \cite{BojanczykMSS2009}, a \myemph{Minsky counter automaton} (short: CA) is essentially a finite state automaton without input but equipped with a finite set of counters which can be incremented, decremented and tested for zero. More formally, a CA $\calM$ is a tuple $(Q, C,\Delta, q_I , F )$, where the set $Q$ of states, the initial state $q_I \in Q$ and the set $F \subseteq Q$ of final states are as in usual finite state automata, and $C$ is a finite set (the \myemph{counters}). The transition relation $\Delta$ is a subset of $$Q \times \{\inc(c), \dec(c), \ifz(c) | c \in C\} \times Q$$ 

A \myemph{configuration} of a CA is a pair $(p, \vec n)$ where $p$ is a state and $\vec n \in \N^C$ gives a value $n_c$ for each counter $c$ in $C$. 
Transitions with $\inc(c)$ can be always applied, whereas transitions with $\dec(c)$ can only be applied to configurations with $n_c > 0$ and 
transitions with $\ifz(c)$ can only be applied if $n_c = 0$. Applying a transition $(p, \inc(c), q)$ to a configuration $(p, \vec n)$ yields 
a configuration $(q, \vec n')$ where $\vec n'$ is obtained from $\vec n$ by incrementing $n_c$ and keeping all other values unchanged. Analogously, applying a (applicable) transition $(p, \dec(c), q)$ to a configuration $(p, \vec n)$ yields a configuration $(q, \vec n')$ where $\vec n'$ is obtained from $\vec n$ by decrementing $n_c$. Applying an (applicable) transition $(p, \ifz(c), q)$ to a configuration $(p, \vec n)$ yields configuration $(q, \vec n)$. A \myemph{run} is a sequence of configurations consistent with $\Delta$. A run is \myemph{accepting}, if it starts at configuration $(q_I, \vec 0)$ and ends in some configuration $(q_F, \vec n)$ with $q_F \in F$. Without loss of generality we require that the counter values in the last configuration of an accepting run are equal to zero.

The emptiness problem for CA is the question whether a given CA has an accepting run. It is well known that the emptiness problem for CA with two counters is undecidable \cite{Minsky1967}.

\proofoflsps

We strongly conjecture that finite satisfiability for the other remaining open case from \cite{ManuelZ13}, namely the extension of two variable logic by one preorder relation and one successor relation of a linear order, is decidable. We are actually working on the details of the proof. However, we felt that the result presented in this note should be made public without further delay.

\bibliography{bibliography}

\end{document}